\documentclass{article}%
\usepackage{amsfonts}
\usepackage{amsmath}
\usepackage{amssymb}
\usepackage{graphicx}%
\providecommand{\U}[1]{\protect\rule{.1in}{.1in}}
\newtheorem{theorem}{Theorem}

\newtheorem{proposition}[theorem]{Proposition}

\begin{document}

\title{Coherent states for compact Lie groups and their large-$N$ limits}
\author{Brian C. Hall\thanks{Supported in part by National Science Foundation grant
DMS-1301534.}\\University of Notre Dame\\Department of Mathematics\\Notre Dame IN 46556, USA\\bhall@nd.edu}
\maketitle

\begin{abstract}
The first two parts of this article surveys results related to the heat-kernel
coherent states for a compact Lie group $K.$ I begin by reviewing the
definition of the coherent states, their resolution of the identity, and the
associated Segal--Bargmann transform. I then describe related results
including connections to geometric quantization and $(1+1)$-dimensional
Yang--Mills theory, the associated coherent states on spheres, and
applications to quantum gravity.

The third part of this article summarizes recent work of mine with Driver and
Kemp on the large-$N$ limit of the Segal--Bargmann transform for the unitary
group $U(N).$ A key result is the identification of the leading-order
large-$N$ behavior of the Laplacian on \textquotedblleft trace
polynomials.\textquotedblright

\end{abstract}
\tableofcontents

\section{Coherent states and Segal--Bargmann transform for Lie groups of
compact type\label{coherent.sec}}

\subsection{Lie groups of compact type and their
complexifications\label{complexify.sec}}

A Lie group $K$ is said to be of compact type if there exists an inner product
on the Lie algebra $\mathfrak{k}$ of $K$ that is invariant under the adjoint
action of $K.$ Compact groups and commutative groups, as well as products of
the two, are of compact type. Conversely, suppose $K$ is a connected Lie group
of compact type and we fix an Ad-$K$-invariant inner product $\left\langle
\cdot,\cdot\right\rangle $ on $\mathfrak{k}.$ Then according to Proposition
2.2 of \cite{Geoquant}, $K$ decomposes as a Lie group direct product
$K=K_{0}\times\mathbb{R}^{k}$ for some $k\geq0,$ where $K_{0}$ is compact and
where the Lie algebras of $K_{0}$ and of $\mathbb{R}^{k}$ are orthogonal with
respect to $\left\langle \cdot,\cdot\right\rangle .$

If $K$ is connected and of compact type, there exists a unique (up to
isomorphism) Lie group $K_{\mathbb{C}}$ with the following properties: (1) the
Lie algebra of $K_{\mathbb{C}}$ is equal to $\mathfrak{k}_{\mathbb{C}%
}:=\mathfrak{k}\oplus i\mathfrak{k},$ (2) $K$ sits inside $K_{\mathbb{C}}$ as
a closed subgroup, and (3) every element $g$ of $K_{\mathbb{C}}$ can be
decomposed uniquely as%
\begin{equation}
g=xe^{iY} \label{polarDecomp}%
\end{equation}
with $x\in K$ and $Y\in\mathfrak{k}.$ We refer to $K_{\mathbb{C}}$ as the
\textbf{complexification} of $K.$ If $K=\mathbb{R}^{k},$ then $K_{\mathbb{C}%
}=\mathbb{C}^{k}$ and if $K$ is the unitary group $U(N),$ then $K_{\mathbb{C}%
}$ is the general linear group $GL(N;\mathbb{C}).$ (For the polar
decomposition in the case $K=U(N),$ see Section 2.5 of \cite{LieBook}.)

We may use the decomposition (\ref{polarDecomp}) to identify the cotangent
bundle $T^{\ast}(K)$ with $K_{\mathbb{C}}$ as follows. We use left-translation
to identify $T^{\ast}(K)$ with $K\times\mathfrak{k}^{\ast}$, then use the
inner product on $\mathfrak{k}$ to identify $K\times\mathfrak{k}^{\ast}$ with
$K\times\mathfrak{k,}$ and finally use the map (\ref{polarDecomp}) to identify
$K\times\mathfrak{k}$ with $K_{\mathbb{C}}.$ In physical terms, we think of
$K$ as the \textit{configuration space} for a physical system and $T^{\ast
}(K)\cong K_{\mathbb{C}}$ as the corresponding \textit{phase space}.

We may consider two physically important examples. First, if $K=SO(3),$ then
$T^{\ast}(K)$ is the configuration space for the rotational degrees of freedom
of a rigid body. Second, the case $K=SU(2)$ plays an important role in
applications to quantum gravity, as described in Section \ref{qg.sec}.

\subsection{Heat kernel}

We fix on the Lie algebra $\mathfrak{k}$ of $K$ an Ad-$K$-invariant inner
product. This inner product determines a bi-invariant Riemannian metric on
$K.$ We let $\Delta_{K}$ denote the associated Laplacian (normalized so that
$\Delta\leq0$). We then let $\rho_{t}$ denote the \textbf{heat kernel} on $K,$
based at the identity. Thus, $\rho_{t}$ satisfies%
\begin{align*}
\frac{d\rho_{t}}{dt}  &  =\frac{1}{2}\Delta_{K}\rho_{t}\\
\lim_{t\rightarrow0^{+}}\rho_{t}  &  =\delta,
\end{align*}
where $\delta$ is a Dirac delta-function at the identity in $K.$ According to
Proposition 1 of \cite{H1}, the quantity $\rho_{t}(x)$ admits an analytic
continuation in the space variable $x$ from $K$ to $K_{\mathbb{C}},$ for each
fixed $t>0.$

\subsection{Coherent states}

We initially take our Hilbert space to the standard \textquotedblleft position
Hilbert space\textquotedblright\ for a particle with configuration space $K,$
namely $L^{2}(K),$ with respect to the Haar measure $dx$ on $K.$ (Later, we
will consider also a Hilbert space of Segal--Bargmann type.) Fix a positive
value $\hbar$ of Planck's constant. For each fixed $g\in K_{\mathbb{C}}$ we
define a \textbf{coherent state }$\chi_{g}\in L^{2}(K)$ by the formula%
\[
\chi_{g}(x)=\overline{\rho_{\hbar}(gx^{-1})},\quad g\in K_{\mathbb{C}}.
\]
Here, since $gx^{-1}$ belongs to $K_{\mathbb{C}},$ the expression $\rho
_{\hbar}(gx^{-1})$ refers to the analytic continuation of the heat kernel in
the space variable. Note that the \textquotedblleft time\textquotedblright%
\ parameter in the heat kernel is now being set equal to Planck's constant.

If $K=\mathbb{R},$ we have $\rho_{t}(x)=(2\pi t)^{-1/2}e^{-x^{2}/(2t)}$ and we
may compute explicitly that for $z=a+ib$ in $K_{\mathbb{C}}=\mathbb{C},$ we
have%
\begin{align*}
\chi_{z}(x)  &  =(2\pi\hbar)^{-1/2}e^{-(\bar{z}-x)^{2}/(2\hbar)}\\
&  =C_{a,b,\hbar}e^{-(a-x)^{2}/(2\hbar)}e^{-ibx/\hbar},
\end{align*}
where $C_{a,b,\hbar}=(2\pi\hbar)^{-1/2}\exp\{(2iab+b^{2})/(2\hbar)\}.$ Thus,
in these cases, the coherent states are the usual Gaussian wave packets, with
$a$ being a position parameter and $b$ being a momentum parameter. (More
precisely, the expected momentum of the coherent state is $-b.$) Note that the
parameter space for the coherent states is the complexified group
$K_{\mathbb{C}},$ which we identify with the phase space $T^{\ast}(K)$ for a
particle moving on $K.$

In the case $K=SU(2),$ which can be identified with the 3-sphere, the coherent
states can be described in terms of the Jacobi theta function. (See Section V
of [HM1].)

\subsection{Resolution of the identity}

Let $\nu_{t}$ be the $K$-invariant heat operator on $K_{\mathbb{C}}.$ This
means, more precisely, that $\nu_{t}$ satisfies the following heat equation
\[
\frac{d\nu_{t}}{dt}=\frac{1}{4}\Delta_{K_{\mathbb{C}}}\nu_{t},
\]
where $\Delta_{K_{\mathbb{C}}}$ is the appropriate left-invariant Laplacian on
$K_{\mathbb{C}}$, subject to the initial condition%
\[
\lim_{t\rightarrow0}\nu_{t}=\delta_{K}.
\]
Here $\delta_{K}$ denotes the Haar measure on $K,$ viewed as a distribution on
$K_{\mathbb{C}}.$ Equivalently, we may think of $\nu_{t}$ as the heat kernel
for the quotient space $K_{\mathbb{C}}/K,$ regarded as a left-$K$-invariant
function on $K_{\mathbb{C}}.$

The coherent states $\chi_{g}$ introduced in the previous subsection then
satisfy the following resolution of the identity:%
\begin{equation}
I=\int_{K_{\mathbb{C}}}\left\vert \chi_{g}\right\rangle \!\left\langle
\chi_{g}\right\vert ~\nu_{t}(g)~dg, \label{resolutionI}%
\end{equation}
where $dg$ is the Haar measure on $K_{\mathbb{C}}.$ The integral in
(\ref{resolutionI}) converges in the weak sense; that is, (\ref{resolutionI})
should be interpreted as meaning that%
\begin{equation}
\left\langle \phi,\psi\right\rangle =\int_{K_{\mathbb{C}}}\left\langle
\phi,\chi_{g}\right\rangle \!\left\langle \chi_{g},\psi\right\rangle ~\nu
_{t}(g)~dg \label{weak}%
\end{equation}
for all $\phi,\psi\in L^{2}(K),$ with absolute convergence of the integral in
(\ref{weak}). (See Theorem 2 in \cite{H1}.)

In the case $K=SU(2),$ the Lie algebra $su(2)$ consists of $2\times2$
skew-Hermitian matrices with trace zero. We may use the Ad-invariant inner
product
\[
\left\langle X,Y\right\rangle =\frac{1}{2}\mathrm{trace}(X^{\ast}Y)
\]
on $su(2).$ (With this choice, the group $SU(2)$ is isometric to the unit
sphere $S^{3}\subset\mathbb{R}^{4}.$) We may then write the resolution of the
identity (\ref{resolutionI}) explicitly, using the polar decomposition
(\ref{polarFormula}), as follows:%
\begin{equation}
I=e^{-\hbar}\int_{SU(2)}\int_{su(2)}\left\vert \chi_{g}\right\rangle
\!\left\langle \chi_{g}\right\vert ~\frac{\sinh(2\left\vert Y\right\vert
)}{2\left\vert Y\right\vert }\frac{e^{-\left\vert Y\right\vert ^{2}/\hbar}%
}{(\pi\hbar)^{3/2}}~dY~dx,\quad g=xe^{iY}.~ \label{explicit}%
\end{equation}
(See Eq. (6) in \cite{HM1} along with the formula for $\nu_{3}$ on p. 1225.)

\subsection{Segal--Bargmann transform}

The \textbf{Segal--Bargmann transform} is a map $C_{\hbar}$ from $L^{2}(K)$
into the space of holomorphic functions on $K_{\mathbb{C}},$ defined by%
\begin{align}
(C_{\hbar}\psi)(g)  &  =\left\langle \chi_{g},\psi\right\rangle \nonumber\\
&  =\int_{K}\rho_{\hbar}(gx^{-1})\psi(x)~dx. \label{cDef}%
\end{align}
According to Theorem 2 of \cite{H1}, we have the following result.

\begin{theorem}
The map $C_{\hbar}$ is a unitary map of $L^{2}(K)$ onto $\mathcal{H}%
L^{2}(K_{\mathbb{C}},\nu_{\hbar}(g)~dg),$ where $\mathcal{H}L^{2}$ denotes the
space of square-integrable holomorphic functions and where $dg$ is the Haar
measure on $K_{\mathbb{C}}.$
\end{theorem}

The fact that $C_{\hbar}\psi$ is holomorphic is equivalent to the fact that
the coherent states $\chi_{g}$ depend antiholomorphically on $g\in
K_{\mathbb{C}}.$ The isometricity of $C_{\hbar},$ meanwhile, is equivalent to
the resolution of the identity (\ref{resolutionI}), in its weak form
(\ref{weak}). The fact that $C_{\hbar}$ maps \textit{onto }$\mathcal{H}%
L^{2}(K_{\mathbb{C}},\nu_{\hbar}(g)~dg),$ however, does not seem to be easily
expressible as a property of the coherent states.

We may interpret $C_{\hbar}\psi$ as a sort of \textit{phase space wave
function} associated to the usual position wave function $\psi.$ If $\psi$ is
a unit vector then the quantity%
\[
\left\vert C_{\hbar}\psi(g)\right\vert ^{2}~dg
\]
is a probability measure on $K_{\mathbb{C}}\cong T^{\ast}(K).$ Results of
\cite{Phasebounds} give sharp upper bounds on the density of this measure
(with respect to the phase volume measure), uniformly over all unit vectors
$\psi.$ This result can be interpreted as a sort of uncertainty principle for
a particle moving on $K,$ that is, as a bound on how concentrated the particle
can be in phase space. In the case $K=\mathbb{R},$ the probability density
$\left\vert C_{\hbar}\psi(g)\right\vert ^{2}$ reduces to the \textit{Husimi
function} associated to $\psi$ (i.e., the convolution of the Wigner function
with a Gaussian smearing function).

There is also inversion formula \cite{Inversion} for the Segal--Bargmann
transform, as follows:%
\[
\psi(x)=\int_{\mathfrak{k}}(C_{\hbar}\psi)(xe^{2iY})\nu_{t/2}(e^{iY})J(Y)~dY,
\]
where $J$ is the Jacobian of the exponential mapping for the quotient space
$K_{\mathbb{C}}/K.$ If we think of $C_{\hbar}\psi$ as a phase space wave
function associated to the position wave function $\psi,$ the inversion
formula says that the position wave function is obtained from the phase space
wave function by integrating out the momentum variables.

\section{Additional results}

\subsection{Geometric quantization\label{geoquant.sec}}

In this section, we discuss a method of obtaining the Segal--Bargmann space
$\mathcal{H}L^{2}(K_{\mathbb{C}},\nu_{\hbar})$ and the associated transform
$C_{\hbar}$ from an apparently completely different perspective, using the
machinery of geometric quantization. (See \cite{Wo} and Chapters 22 and 23 of
\cite{QuantumBook} for general information about geometric quantization.)

To perform geometric quantization on a symplectic manifold $(M,\omega)$, we
first construct a \textit{prequantum line bundle} $L$ over our phase space,
together with a Hermitian structure and a connection $\nabla$ on $L,$ where
the curvature of $\nabla$ is equal to $\omega/\hbar.$ (Such a
line-bundle-with-connection exists provided that the integral of $\omega
/(2\pi\hbar)$ over every closed surface $S$ in $M$ is an integer.) We then
construct a \textit{polarization} on $L,$ which means, roughly, a choice of a
half-dimensional set of directions at each point in the phase space. The
quantum Hilbert space then consists of the space of square-integrable sections
of $L$ that are \textquotedblleft polarized,\textquotedblright\ that is, those
that are covariantly constant in the directions of the polarization. If the
polarization is purely real, the Hilbert space will be something like the
usual position Hilbert space, while if the polarization is purely complex, the
Hilbert space will be something like the Segal--Bargmann space.

One important additional aspect of geometric quantization is the
\textit{half-form correction}, also referred to as the metaplectic correction.
(See Sections 23.6 and 23.7 of \cite{QuantumBook}.) This correction is needed
in the case of real polarizations to obtain a natural inner product on the
space of polarized sections. In the case of complex polarizations, the
half-form correction is not strictly necessary, but often leads to better
results. As an example, if we quantize the harmonic oscillator by means of a
complex polarization on the plane that is invariant under the classical
dynamics, the half-form correction leads to the \textquotedblleft
correct\textquotedblright\ energy levels of the Hamiltonian, $\hbar
\omega(n+1/2),$ with the $1/2$ coming from the half-forms. (See \cite[Example
23.53]{QuantumBook}.)

In the case at hand, we take our phase space to be the cotangent bundle
$T^{\ast}(K),$ with $\omega$ being the canonical 2-form, given in coordinates
as $\omega=\sum dp_{j}\wedge dx_{j}$. Let $\theta$ be the canonical 1-form,
given in coordinates as $\sum p_{j}~dx_{j},$ so that $d\theta=\omega.$ Then we
may take $L$ to be the trivial bundle with trivial Hermitian structure and
connection $\nabla$ given by $\nabla_{X}=X-(i/\hbar)\theta(X).$ We then
construct a polarization by means of the identification of $T^{\ast}(K)$ with
$K_{\mathbb{C}},$ discussed in Section \ref{complexify.sec}. Thanks to work of
Guillemin and Stenzel \cite{GS1,GS2}, we know that the function%
\[
\kappa(x,Y)=\frac{1}{2}\left\vert Y\right\vert ^{2}%
\]
is a K\"{a}hler potential. This implies that the function%
\[
s_{0}(Y)=e^{-\left\vert Y\right\vert ^{2}/(2\hbar)}%
\]
is a polarized section of $L$. A general polarized section then has the form%
\[
Fs_{0},
\]
where $F$ is a holomorphic function on $K_{\mathbb{C}}\cong T^{\ast}(K).$

The \textit{canonical bundle} $\kappa$ associated to the given polarization is
now the bundle whose sections $(n,0)$ forms on $K_{\mathbb{C}},$ where $n$ is
the complex dimension of $K_{\mathbb{C}}.$ This bundle is trivial and we may
choose a nowhere-vanishing, bi-invariant holomorphic section $\alpha.$ We may
then construct a trivial square root $\delta$ of the canonical bundle with a
trivializing section $\sqrt{\alpha}.$ Elements of the half-form corrected
quantum Hilbert space are then polarized sections of $L\otimes\delta.$
Explicitly, these have the form%
\begin{equation}
s=Fe^{-\left\vert Y\right\vert ^{2}/(2\hbar)}\otimes\sqrt{\alpha},
\label{sForm}%
\end{equation}
where $F$ is a holomorphic function on $K_{\mathbb{C}}\cong T^{\ast}(K).$

To compute the norm of such a section, we must understand how to compute the
pointwise magnitude of $\sqrt{\alpha}.$ To do this, we square $\sqrt{\alpha}$
to get $\alpha,$ then wedge the result with its complex conjugate, to get a
$2n$-form. We then compare this $2n$-form to the Liouville volume form
$\omega^{n}/n!.$ Finally, we take a square root, so that the resulting
expression is quadratic in $\sqrt{\alpha}$:%
\[
\left\vert \sqrt{\alpha}\right\vert ^{2}=\left[  c\frac{\alpha\wedge
\bar{\alpha}}{\omega^{n}/n!}\right]  ^{1/2}.
\]
Here $c$ is a universal constant chosen so that $c(\alpha\wedge\bar{\alpha})$
is a positive multiple of the Liouville form. We then define the norm of the
section $s$ in (\ref{sForm}) as%
\[
\left\Vert s\right\Vert ^{2}=\int_{T^{\ast}(K)}\left\vert F\right\vert
^{2}e^{-\left\vert Y\right\vert ^{2}/\hbar}\left\vert \sqrt{\alpha}\right\vert
^{2}~\frac{\omega^{n}}{n!}.
\]

\begin{theorem}
Under the identification (\ref{polarDecomp}) of $T^{\ast}(K)$ with
$K_{\mathbb{C}},$ the measure
\[
e^{-\left\vert Y\right\vert ^{2}/\hbar}\left\vert \sqrt{\alpha}\right\vert
^{2}~\frac{\omega^{n}}{n!}%
\]
on $T^{\ast}(K)$ coincides up to a constant $c_{\hbar}$ with the $K$-invariant
heat kernel measure $\nu_{\hbar}(g)~dg$ on $K_{\mathbb{C}}.$ Thus, the
half-form corrected quantum Hilbert space may be identified naturally with the
Segal--Bargmann space over $K_{\mathbb{C}},$ namely $\mathcal{H}%
L^{2}(K_{\mathbb{C}},\nu_{\hbar}).$
\end{theorem}

This result is Theorem 2.5 in \cite{Geoquant}. The result is surprising in
that it is not obvious how geometric quantization \textquotedblleft
knows\textquotedblright\ about the heat kernel. The agreement between the
results of geometric quantization and the seemingly unrelated heat-kernel
methods described in Section \ref{coherent.sec} suggests that there is
something \textquotedblleft right\textquotedblright\ about the space
$\mathcal{H}L^{2}(K_{\mathbb{C}},\nu_{\hbar}).$

We have also a result that relates the BKS pairing map of geometric
quantization (e.g., Section 23.8 of \cite{QuantumBook}) to the Segal--Bargmann transform.

\begin{theorem}
The pairing map on $T^{\ast}(K)$ between the vertically polarized space and
the K\"{a}hler-polarized space is a constant multiple of the Segal--Bargmann
transform $C_{\hbar}.$
\end{theorem}

This result is Theorem 2.6 in \cite{Geoquant}.

Various additional works have shed light on the preceding results. In
particular, work of Florentino, Matias, Mour\~{a}o, and Nunes
\cite{FMMN1,FMMN2} and then of Lempert and Sz\H{o}ke \cite{LS1,LS2,Sz}
consider a \textit{family} of complex structures on $T^{\ast}(K).$ (The family
is described by one real parameter in the case of \cite{FMMN1,FMMN2} and two
real parameters in the case of \cite{LS1,LS2,Sz}.) For each complex structure,
one can perform geometric quantization to get a Hilbert space. These Hilbert
spaces form a \textquotedblleft field\textquotedblright\ of Hilbert spaces
over the the parameter space, which in some cases is actually a
\textquotedblleft Hilbert bundle.\textquotedblright\ One can then consider a
connection on this field and use parallel transport to identify different
spaces. The resulting identifications are closely related to the
Segal--Bargmann transform $C_{\hbar},$ thus giving a more geometric
perspective on results of \cite{Geoquant}.

\subsection{$(1+1)$-dimensional Yang--Mills theory}

We now describe results linking the heat kernel coherent states $\chi_{g}$ on
a compact Lie group with the canonical quantization of $(1+1)$-dimensional
Yang--Mills theory. The first results in this direction were obtained by
Landsman and Wren \cite{LW} (in the commutative case) and by Wren \cite{Wr}
(in the general case). Similar results using a different approach were then
obtained by Driver and the author in \cite{DH1}; see also the expository paper
\cite{RevMathPhys}.

We consider canonical quantization of $(1+1)$-dimensional Yang--Mills theory
on a space-time cylinder $S^{1}\times\mathbb{R},$ with structure group $K.$ If
we work in the temporal gauge, the configuration space for the theory is the
space $\mathcal{A}$ of connections on the spatial circle. We consider the
\textbf{gauge group} $\mathcal{G},$ consisting of gauge transformations that
preserve the temporal gauge, namely the group of maps of $S^{1}$ into $K.$ We
consider also the \textbf{based gauge group} $\mathcal{G}_{0}$ consisting of
maps of $S^{1}$ into $K$ that are equal to the identity at one fixed point in
the circle. Restricting attention at first to the based gauge group simplifies
the analysis, because $\mathcal{G}_{0}$ acts freely on $\mathcal{A}$, so that
the quotient is a manifold (in this case, finite dimensional). The quotient of
$\mathcal{A}$ by $\mathcal{G}_{0}$ is naturally identified with the structure
group $K.$ This identification is easy to understand: The holonomy of a
connection around the spatial circle is (fully) invariant under the action of
$\mathcal{G}_{0}$ and in the circle case, this is the only gauge-invariant quantity.

The phase space for the unreduced system is the cotangent bundle $T^{\ast
}(\mathcal{A}),$ which may also be identified with $\mathcal{A}_{\mathbb{C}},$
the space of complex connections. The reduced phase space is the symplectic
quotient of $T^{\ast}(\mathcal{A})$ by the action of $\mathcal{G}_{0},$ which
is constructed by first restricting to a constraint set (the zero set of the
momentum map) and then quotienting by $\mathcal{G}_{0}.$ This symplectic
quotient may be identified either with $T^{\ast}(K)$ or with $K_{\mathbb{C}}.$

One can then attempt to a Segal--Bargmann-type quantization of the phase space
$\mathcal{A}_{\mathbb{C}}.$ Since $\mathcal{A}_{\mathbb{C}}$ is just a vector
space, this is not difficult to do rigorously. The difficulty comes when one
tries to impose the quantum version of gauge symmetry: There are no nonzero,
gauge-invariant states that have finite norm \cite{DH2}. To work around this
problem, one must perform some sort of \textquotedblleft gauge
fixing\textquotedblright\ (which does not necessarily involve choosing one
element out of each gauge orbit). This has been done in two different ways.
First, Wren \cite{Wr}, using integration over the gauge group, develops a
method of \textquotedblleft projecting\textquotedblright\ the coherent states
for $\mathcal{A}_{\mathbb{C}}$ onto the (nonexistent) gauge-invariant
subspace. Second, Driver and the author use the Segal--Bargmann space over
$\mathcal{A}_{\mathbb{C}}$ with a large-variance Gaussian measure that
approximates the nonexistent Lebesgue measure. (See \cite{DH1} as well as the
expository paper \cite{RevMathPhys}.)

Both approaches give the same conclusion: If one takes the coherent states for
the infinite-dimensional linear space $\mathcal{A}_{\mathbb{C}}$ and projects
them on to the gauge-invariant subspace, the resulting states are precisely
the heat-kernel coherent states $\chi_{g}$ for the reduced phase space
$T^{\ast}(K)\cong K_{\mathbb{C}}.$ As with the results concerning geometric
quantization in Section \ref{geoquant.sec}, it is reassuring to see the same
coherent states arise from a method seemingly unrelated to those in Section
\ref{coherent.sec}. In particular, the heat kernel again arises here in a
natural way, without having been put in by hand.

This result, together with the results of the previous subsection can be
interpreted as an instance of the notion of \textquotedblleft quantization
commutes with reduction.\textquotedblright\ More specifically, this is an
instance in which quantization commutes \textit{unitarily} with reduction, as
we now explain. In the setting of holomorphic (or K\"{a}hler) quantization, an
influential paper of Guillemin and Sternberg \cite{GuilleminSternberg} showed
that there is a natural vector space isomorphism between two spaces: On the
one hand, the Hilbert space obtained by \textit{first} quantizing and
\textit{then} reducing by the action of a compact group, and on the other
hand, the Hilbert space obtained by doing these operations in the other order.

Guillemin and Sternberg did not, however, establish any sort of unitary result
for their map. Indeed, results of \cite{HallKirwin} show that the
Guillemin--Sternberg map is \textit{not even asymptotically unitary} as
$\hbar$ tends to zero. Nevertheless, \cite{HallKirwin} shows that if one
includes half-forms in the quantization, one obtains asymptotic unitarity. It
still remains to look for examples where the map is not just asymptotically
unitary, but unitary on the nose. By combining the results of \cite{DH1} and
\cite{Geoquant}, we obtain an instance of exact unitarity. If we quantize
$\mathcal{A}_{\mathbb{C}}$ first and then pass (after a gauge fixing) to the
gauge-invariant subspace, we obtain \textit{the same Hilbert space with the
same inner product} as if we first reduce $\mathcal{A}_{\mathbb{C}}$ by
$\mathcal{G}_{0}$ and then quantize with half-forms. (Compare Section 8 of
\cite{RevMathPhys}.)

\subsection{Coherent states on spheres}

The results of \cite{H1} extend in an obvious way to \textit{normal
homogeneous spaces}, that is, manifolds of the form $K/H$ where $H$ is a
closed subgroup of $K$ and where the metric on $K/H$ is induced in a natural
way from the metric on $K.$ In the case that $K/H$ is a \textit{symmetric
space} (i.e., when $H$ is the fixed-point subgroup of an involution of $K$),
results of Stenzel \cite{St} give a much nicer description of the resulting
Segal--Bargmann space and transform than the one in \cite{H1}. In particular,
Stenzel describes the space and the inverse transform in terms of the heat
kernel on the dual noncompact symmetric space to $K/H.$

The results of \cite{H1} and \cite{St} apply, in particular, to the case of an
$n$-sphere $S^{n}=SO(n+1)/SO(n).$ In this case, the dual noncompact symmetric
space is $n$-dimensional hyperbolic space. We emphasize that the sphere
$S^{n}$ is playing the role of the \textit{configuration space} of a classical
system, with the phase space being $T^{\ast}(S^{n}).$ (Thus, the results
discussed here are essentially unrelated to the study of coherent states on a
2-sphere, viewed as the phase space of a classical system.) This special case
has received special attention because of its simplicity and physical applications.

In \cite{KR1}, Kowalski and Rembieli\'{n}ski independently introduced the same
heat-kernel coherent states as in \cite{H1} and \cite{St}, but from a
different point of view, using a polar decomposition method. (See also
\cite{KR2}.) Meanwhile, Thiemann in \cite{Thie1} proposed a general
\textquotedblleft complexifier\textquotedblright\ method of constructing
coherent states and Segal--Bargmann-type transforms. The author and Mitchell
in \cite{HM1} then examined the sphere case in detail, incorporating both the
polar-decomposition method and the complexifier method. The article \cite{HM2}
then examines the large-radius limit (in the odd-dimensional case), showing
that the coherent states converge in this limit to the usual Gaussian wave
packets on $\mathbb{R}^{n}.$ The article \cite{HM3} then considers the case of
a particle moving on a 2-sphere in the presence of a constant magnetic field.
Finally, Kowalski, Rembieli\'{n}ski, and Zawadzki in \cite{KRZ} examine
numerically the free dynamics of coherent states on $S^{2}.$

\subsection{Applications to quantum gravity\label{qg.sec}}

The coherent states $\chi_{g}$ for compact Lie groups, especially in the
$SU(2)$ case, have been used extensively in the literature on loop quantum
gravity. The first such application was in the paper \cite{ALMMT} of Ashtekar,
Lewandowski, Marolf, Mour\~{a}o, and Thiemann. These authors work in
Ashtekar's \textquotedblleft new variables\textquotedblright\ for gravity and
construct a Segal--Bargmann-type transform designed to deal with the reality
conditions in the original, complex-valued version of the Ashtekar variables.
Since then, work of Thiemann and others have brought a real-valued version of
the Ashtekar variables to the fore. In this setting, the coherent states serve
mainly as a tool for investigating semiclassical properties of loop quantum
gravity. This perspective was developed first in a series of papers by
Thiemann \cite{Thie2}, by Thiemann and Winkler \cite{TW}, and by Bahr and
Thiemann \cite{BT}. Since then the coherent states of \cite{H1} have continued
to be widely used in quantum gravity, with too many papers to cite individually.

\section{The large-$N$ limit}

In this section, we describe work on the large-$N$ limit of the
Segal--Bargmann transform for the unitary group $U(N).$ One motivation for
letting $N$ tend to infinity comes from the literature on quantum field
theory, where limits of this sort are popular in various sorts of gauge
theories. The idea appears to have originated with 't Hooft \cite{tHooft}, who
suggested that $U(N)$ Yang--Mills theory simplifies in the large-$N$ limit,
with the usual path-integral concentrating onto a single connection known as
the \textquotedblleft master field.\textquotedblright\ Meanwhile, work of
Gross and Taylor \cite{GT1} on the large-$N$ limit of two-dimensional
Yang--Mills theory shows a connection with string theory. We mention, finally,
the paper \cite{Mal} of Maldacena on the large-$N$ limit of superconformal
field theories and its connection to supergravity, which has over 4,000
citations in the Science Citation Index. The methods used here are closely
related to those used in the study of the large-$N$ limit of Yang--Mills
theory on the plane, as in \cite{levy} or \cite{DHK2}.

Another motivation for letting $N$ tend to infinity comes from random matrix
theory, in which the structure of the eigenvalues of random matrices
simplifies as the size of the matrices goes to infinity. This subject began
with work of Wigner in the 1950's in nuclear physics, but has now blossomed
into a discipline unto itself. To see something of the connection with random
matrix theory, note that we will consider a probability measure on $U(N)$,
given by the heat kernel $\rho_{t}.$ In the large-$N$ limit, the distribution
of the eigenvalues of random matrices selected according to $\rho_{t}$ have a
deterministic limit, identified by Biane. This limit should be understood as a
deformation of the classical Wigner semicircular distribution.

\subsection{Overview of large-$N$ limit}

In this subsection, we give an overview of results on the large-$N$ limit of
the Segal--Bargmann transform on $U(N)$; more details will be provided in the
subsequent subsections. The results given here are based on joint work with
Driver and Kemp \cite{DHK}, which in turn was motivated by earlier work of
Biane \cite{Biane2}. Results similar to those in \cite{DHK} were obtained
independently by C\'{e}bron in \cite{Ceb}. A more detailed exposition of some
of these results may be found in the author's unpublished preprint
\cite{Exposit}.

Throughout the rest of the paper, we restrict our attention to the group
$K=U(N)$ (the group of $n\times n$ unitary matrices) and its complexification,
$K_{\mathbb{C}}=GL(N;\mathbb{C})$ (the group of all $n\times n$ invertible
matrices). We also use slightly different notation, for consistency with
\cite{DHK}. Notably, we revert to using $t$ for the time-parameter in the
relevant heat kernels, rather than $\hbar$ as in the previous sections.

We use on $U(N)$ the bi-invariant metric whose value on the Lie algebra $u(N)$
of $U(N)$ is given by the \textit{scaled} Hilbert--Schmidt inner product,%
\[
\left\langle X,Y\right\rangle _{N}:=N\mathrm{Trace}(X^{\ast}Y).
\]
The motivation for this scaling is described in the next subsection. The above
inner product gives rise to a bi-invariant metric on $U(N)$ and then to a
bi-invariant Laplacian $\Delta_{N}.$

We consider now the \textquotedblleft B-version\textquotedblright%
\ Segal--Bargmann transform in Theorem 1$^{\prime}$ of \cite{H1}, which has
better large-$N$ behavior than the \textquotedblleft
C-version\textquotedblright\ transform considered previously. (In \cite{DHK},
we actually consider a two-parameter version of the Segal--Bargmann transform,
which includes the B-version as a special case.) For each~$t>0,$ the transform
$B_{t}^{N}$ is defined by the same formula as $C_{t}$:
\[
(B_{t}^{N}f)(g)=\int_{U(N)}\rho_{t}(gx^{-1})f(x)~dx,
\]
where $\rho_{t}$ is the heat kernel on $U(N)$ with respect to the metric
coming from the scaled inner product on (\ref{unInner}). The difference
between the $B_{t}$ and $C_{t}$ transforms is that we use different norms. For
$B_{t},$ we take use the $L^{2}$ norm with respect to the heat kernel measure
$\rho_{t}(x)~dx.$ On the range side, we use the $L^{2}$ norm with respect to
the measure $\mu_{t}(g)~dg,$ where $\mu_{t}$ is the \textquotedblleft
full\textquotedblright\ heat kernel for $GL(N;\mathbb{C}),$ that is, the one
that concentrates to a $\delta$-function at the identity as $t\rightarrow0.$
(Recall that the measure $\nu_{t}$ concentrates to the $\delta$-measure on $K$
as $t\rightarrow0.$)

\begin{theorem}
For each $N>0,$ the transform $B_{t}^{N}$ is unitary from $L^{2}(U(N),\rho
_{t})$ onto $\mathcal{H}L^{2}(GL(N;\mathbb{C}),\mu_{t}).$
\end{theorem}

We may extend the transform to act on functions on $U(N)$ with values in
$M_{N}(\mathbb{C}),$ space of all $N\times N$ matrices with complex entries.
The extension is accomplished by applying the scalar transform
\textquotedblleft entrywise.\textquotedblright\ We denote the resulting
\textbf{boosted\ Segal--Bargmann transform} by $\mathbf{B}_{t}^{N}$:%
\[
\mathbf{B}_{t}^{N}:L^{2}(U(N),\rho_{t}^{N};M_{N}(\mathbb{C}))\rightarrow
\mathcal{H}L^{2}(GL(N;\mathbb{C}),\mu_{t}^{N};M_{N}(\mathbb{C})).
\]
As proposed by Biane in \cite{Biane2}, we apply $\mathbf{B}_{t}^{N}$ to
\textbf{single-variable polynomial functions} on $U(N)$\ that is, functions of
the form%
\begin{equation}
f(U)=c_{0}I+c_{1}U+c_{2}U^{2}+\cdots+c_{N}U^{N},\quad U\in U(N), \label{polyU}%
\end{equation}
where $c_{0},\ldots,c_{N}$ are constants.

If we apply $\mathbf{B}_{t}^{N}$ to such a polynomial function, the result
will typically \textit{not} be a polynomial function on $GL(N;\mathbb{C}).$
Rather, the result will be a \textbf{trace polynomial function}\ on
$GL(N;\mathbb{C}),$ that is, a linear combination of functions of the form%
\begin{equation}
Z^{k}\mathrm{tr}(Z)\mathrm{tr}(Z^{2})\cdots\mathrm{tr}(Z^{M}),\quad Z\in
GL(N;\mathbb{C}), \label{tracePoly}%
\end{equation}
where $k$ and $M$ are non-negative integers. Here $\mathrm{tr}(\cdot)$ is the
\textbf{normalized trace} given by%
\begin{equation}
\mathrm{tr}(A)=\frac{1}{N}\sum_{j=1}^{N}A_{jj} \label{normalizedTrace}%
\end{equation}
for any $A\in M_{N}(\mathbb{C}).$

Although for any one fixed value of $N,$ the boosted transform $\mathbf{B}%
_{t}^{N}$ does not map polynomial functions on $U(N)$ to polynomial functions
on $GL(N;\mathbb{C}),$ there is a sense in which \textbf{the large-}%
$N$\textbf{\ limit} of $\mathbf{B}_{t}^{N}$ does have this property. To
understand how this works, let consider the example of the matrix-valued
function
\[
f(U)=U^{2}%
\]
on $U(N).$ Then, according to Example 3.5 of \cite{DHK}, we have%
\begin{equation}
\mathbf{B}_{t}^{N}(f)(Z)=e^{-t}\left[  \cosh(t/N)Z^{2}-t\frac{\sinh(t/N)}%
{t/N}Z\mathrm{tr}(Z)\right]  ,\quad Z\in GL(N;\mathbb{C}). \label{btUsquared1}%
\end{equation}

If we formally let $N$ tend to infinity in (\ref{btUsquared1}), we obtain%
\begin{equation}
\lim_{N\rightarrow\infty}\mathbf{B}_{t}^{N}(f)(Z)=e^{-t}[Z^{2}-tZ\mathrm{tr}%
(Z)]. \label{btUsquared2}%
\end{equation}
The right-hand side of (\ref{btUsquared2}) is, apparently, still a trace
polynomial and not a single-variable polynomial as in (\ref{polyU}). There is,
however, another limiting phenomenon that occurs when $N$ tends to infinity,
in addition to the convergence of the coefficients of $Z^{2}$ and
$Z\mathrm{tr}(Z)$ in (\ref{btUsquared1}), namely, the phenomenon of
\textbf{concentration of trace}.

As $N$ tends to infinity, the function $\mathrm{tr}(U^{k})$ in $L^{2}%
(U(N),\rho_{t}^{N})$ converges as $N$ tends to infinity to a certain constant
$\nu_{k}(t),$ in the sense that%
\[
\lim_{N\rightarrow\infty}\left\Vert \mathrm{tr}(U^{k})-\nu_{k}(t)\right\Vert
_{L^{2}(U(N),\rho_{t}^{N})}=0.
\]
What this means, more accurately, is that the \textit{measure} $\rho_{t}^{N}$
on $U(N)$ is concentrating, as $N$ tends to infinity with $t$ fixed, onto the
set where~$\mathrm{tr}(U^{k})=\nu_{k}(t).$ A similar concentration of trace
phenomenon occurs in $GL(N;\mathbb{C}),$ except that in this case, all of the
traces concentrate to the value 1:%
\[
\lim_{N\rightarrow\infty}\left\Vert \mathrm{tr}(Z^{k})-1\right\Vert
_{L^{2}(GL(N;\mathbb{C}),\mu_{t}^{N})}=0.
\]

Thus, the \textquotedblleft correct\textquotedblright\ way to evaluate the
large-$N$ limit in (\ref{btUsquared1}) is in two stages. First, we take the
limit as $N$ tends to infinity of the coefficients of $Z^{2}$ and
$Z\mathrm{tr}(Z),$ as in (\ref{btUsquared2}). Second, we replace
$\mathrm{tr}(Z)$ by the constant 1. The result is%
\begin{equation}
\lim_{N\rightarrow\infty}\mathbf{B}_{t}^{N}(f)(Z)=e^{-t}[Z^{2}-tZ].
\label{btUsquared3}%
\end{equation}
Note that the right-hand side of (\ref{btUsquared3}) is, for each fixed value
of $t,$ a single-variable polynomial in $Z.$

In \cite{DHK}, we show that a similar phenomenon occurs in general. Given any
polynomial $p$ in a single variable, let $p_{N}$ denote the matrix-valued
function on $U(N)$ obtained by plugging a variable $U\in U(N)$ into $p,$ as in
(\ref{polyU}). We also allow $p_{N}$ to denote the similarly defined function
on $GL(N;\mathbb{C}).$

\begin{theorem}
[Driver--Hall--Kemp]\label{mainIntro.thm}Let $p$ be a polynomial in a single
variable. Then for each fixed $t>0,$ there exists a unique polynomial $q_{t}$
in a single variable such that%
\begin{equation}
\lim_{N\rightarrow\infty}\left\Vert \mathbf{B}_{t}^{N}(p_{N})-(q_{t}%
)_{N}\right\Vert _{L^{2}(GL(N;\mathbb{C}),\mu_{t}^{N};M_{N}(\mathbb{C}))}=0.
\label{introLim}%
\end{equation}

\end{theorem}

If, for example, $p$ is the polynomial $p(u)=u^{2},$ then $q_{t}$ is the
polynomial given by%
\[
q_{t}(z)=e^{-t}(z^{2}-tz),
\]
so that
\[
(q_{t})_{N}(Z)=e^{-t}(Z^{2}-tZ),\quad Z\in GL(N;\mathbb{C}),
\]
as on the right-hand side of (\ref{btUsquared3}).

In \cite{DHK}, we also show that the map $p\mapsto q_{t}$ coincides with the
\textquotedblleft free Hall transform\textquotedblright\ of Biane, denoted
$\mathcal{G}^{t}$ in \cite{Biane2}. Although it was conjectured in
\cite{Biane2} that $\mathcal{G}^{t}$ is the large-$N$ limit of $\mathbf{B}%
_{t}^{N}$ as in (\ref{introLim}), Biane actually constructs $\mathcal{G}^{t}$
by using free probability. Theorem \ref{mainIntro.thm} was also proved
independently by G. Cebr\'{o}n \cite{Ceb}, using substantially different
methods. Besides using different methods from \cite{Ceb}, the paper \cite{DHK}
establishes a \textquotedblleft two-parameter\textquotedblright\ version of
Theorem \ref{mainIntro.thm}.

A key tool in proving the results described above is the \textbf{asymptotic
product rule} for the Laplacian on $U(N).$ This rule states that---on certain
classes of functions and for large values of $N$---the Laplacian behaves like
a \textit{first-order} differential operator. That is to say, in the usual
product rule for the Laplacian, the cross terms are small compared to the
other two terms. The asymptotic product rule provides the explanation for the
concentration of trace phenomenon and is also the key tool we use in deriving
a recursive formula for the polynomials $q_{t}$ in Theorem \ref{mainIntro.thm}.

\subsection{The Laplacian and Segal--Bargmann transform on $U(N)$}

In the rest of the article, we provide more details on the results presented
in the preceding subsection. We consider $U(N),$ the group of $N\times N$
unitary matrices. The Lie algebra $u(N)$ of $U(N)$ is the $N^{2}$-dimensional
real vector space consisting of $N\times N$ matrices $X$ with $X^{\ast}=-X.$
We use on $u(N)$ the following Ad-invariant inner product $\left\langle
\cdot,\cdot\right\rangle _{N}$:%
\begin{equation}
\left\langle X,Y\right\rangle _{N}=N\mathrm{Trace}(X^{\ast}Y), \label{unInner}%
\end{equation}
where $\mathrm{Trace}$ is the ordinary trace, $\mathrm{Trace}(A)=\sum
_{j}A_{jj}.$ (This inner product is real valued for $X,Y\in u(N).$) This inner
product on $u(N)$ determines a bi-invariant Riemannian metric on $U(N),$ which
in turn determines a Laplace operator $\Delta_{N}.$ Note that $u(N)$ is the
space of skew-Hermitian matrices, which may be identified with the Hermitian
matrices by means of the map $X\mapsto iX.$ The Gaussian measure
\[
Ce^{-\left\langle X,X\right\rangle _{N}/2}~dX
\]
on $u(N)\cong\{\mathrm{Hermitian~matrices}\}$ is then one commonly called the
\textit{Gaussian unitary ensemble} in random matrix theory. This observation
gives one motivation for the particular scaling used in (\ref{unInner}).

The following example will given another motivation for the scaling by a
factor of $N$ in (\ref{unInner}). Consider the action of $\Delta_{N}$ on the
matrix entries for the standard representation of $U(N),$ that is, functions
of the form $f_{jk}(U)=U_{jk}.$ It follows from the $k=1$ case of Proposition
\ref{lapPower.prop} below that%
\begin{equation}
\Delta_{N}(U_{jk})=-U_{jk}. \label{eigen}%
\end{equation}
That is, the functions $f_{jk}$ are eigenvalues for $\Delta_{N}$ with
eigenvalue $-1,$ for all $N$ and all $j,k.$ In particular, the normalization
of the inner product in (\ref{unInner}) has the result that the eigenvalues of
$\Delta_{N}$ in the standard representation are \textit{independent of }$N.$
By contrast, if we had omitted the factor of $N$ in (\ref{unInner}), we would
have had $\Delta_{N}(U_{jk})=-NU_{jk},$ which would not bode well for trying
to take the $N\rightarrow\infty$ limit. (Note that the inner product and the
Laplacian scale oppositely; the factor of $N$ in (\ref{unInner}) produces a
factor of $1/N$ in the formula for $\Delta_{N},$ which scales the eigenvalues
from $-N$ to $-1.$)

Our goal is now to understand the behavior of $B_{t}^{N}$ as $N$ tends to
infinity. As the preceding discussion suggests, for this limit to have a
chance to exist, the factor of $N$ scaling in (\ref{unInner}) is essential.
Indeed, results of Gordina \cite[Sect. 8]{Go1} show that if we used the
unscaled Hilbert--Schmidt inner product on the Lie algebra, we would not
obtain meaningful transform in the limit.

For reasons that will be explained later, it is desirable to extend the
transform $B_{t}^{N}$ to a \textquotedblleft boosted\textquotedblright%
\ transform $\mathbf{B}_{t}^{N},$ acting on matrix valued functions as
follows. Given $f:U(N)\rightarrow M_{N}(\mathbb{C}),$ we apply the scalar
transform $B_{t}^{N}$ \textquotedblleft entrywise.\textquotedblright\ That is,
$\mathbf{B}_{t}^{N}f$ is the holomorphic function $F:GL(N;\mathbb{C}%
)\rightarrow M_{N}(\mathbb{C})$ whose $(j,k)$ entry is $B_{t}^{N}(f_{jk}).$ We
define the norm of matrix-valued functions on $U(N)$ or $GL(N;\mathbb{C})$ as
follows:
\begin{align}
\left\Vert f\right\Vert _{L^{2}(U(N),\rho_{t}^{N};M_{N}(\mathbb{C}))}^{2}  &
=\int_{U(N)}\mathrm{tr}(f(U)^{\ast}f(U))~d\rho_{t}^{N}(U)\label{MnNorm}\\
\left\Vert f\right\Vert _{L^{2}(GL(N;\mathbb{C}),\mu_{t}^{N};M_{N}%
(\mathbb{C}))}^{2}  &  =\int_{GL(N;\mathbb{C})}\mathrm{tr}(f(Z)^{\ast
}f(Z))~d\mu_{t}^{N}(Z), \label{MnNorm2}%
\end{align}
where $\mathrm{tr}(\cdot)$ is the \textit{normalized} trace defined in
(\ref{normalizedTrace}). Note that the normalization of the Hilbert--Schmidt
norm in (\ref{MnNorm}) and (\ref{MnNorm2}) is different from the one we use in
(\ref{unInner}) to define the Laplacian $\Delta_{N}.$ The normalizations in
(\ref{MnNorm}) and (\ref{MnNorm2}) ensure that in both Hilbert spaces, the
constant function $f(U)=I$ has norm one.

\subsection{The action of the Laplacian on trace polynomials}

We will be interested in the action of $\Delta_{N}$ on \textbf{trace
polynomials}, that is, on matrix-valued functions that are linear combinations
of functions of the form%
\begin{equation}
U^{k}\mathrm{tr}(U)\mathrm{tr}(U^{2})\cdots\mathrm{tr}(U^{n})
\label{tracePoly1}%
\end{equation}
for some $k$ and $n.$ (More generally, we could consider a more generally
trace \textit{Laurent }polynomials, where we allow negative powers of $U$ and
traces thereof.) The formula the action of $\Delta_{N}$ on such functions was
originally worked out by Sengupta; see Definition 4.2 and Lemma 4.3 in
\cite{Sengupta}. We begin by recording the formula for the Laplacian of a
single power of $U.$

\begin{proposition}
\label{lapPower.prop}For each positive integer $k,$ we have%
\begin{equation}
\Delta_{N}(U^{k})=-kU^{k}-2\sum_{m=1}^{k-1}mU^{m}\mathrm{tr}(U^{k-m}),
\label{lapPower1}%
\end{equation}
and%
\begin{equation}
\Delta_{N}(\mathrm{tr}(U^{k}))=-k\mathrm{tr}(U^{k})-2\sum_{m=1}^{k-1}%
m\mathrm{tr}(U^{m})\mathrm{tr}(U^{k-m}). \label{lapPower2}%
\end{equation}

\end{proposition}

This result is Theorem 3.3 in \cite{DHK}. Note that when $k=1,$ the sums on
the right-hand sides of (\ref{lapPower1}) and (\ref{lapPower2}) are empty.
Thus, actually, $\Delta_{N}(U)=-U$ and $\Delta_{N}(\mathrm{tr}%
(U))=-\mathrm{tr}(U).$ Since, by definition, $\Delta_{N}$ acts
\textquotedblleft entrywise\textquotedblright\ on matrix-valued functions, the
assertion that $\Delta_{N}(U)=-U$ is equivalent to the assertion that
$\Delta_{N}(U_{jk})=-U_{jk}$ for all $j$ and $k.$ An elementary proof of
Proposition \ref{lapPower.prop} is outlined in Section 9 of \cite{Exposit}.

Let us make a few observations about the formulas in Proposition
\ref{lapPower.prop}. First, since we are supposed to be considering
matrix-valued functions, we should really think of $\mathrm{tr}(U^{k})$ as the
matrix-valued function $U\mapsto\mathrm{tr}(U^{k})I.$ Nevertheless, if we
chose to think of $\mathrm{tr}(U^{k})$ as a scalar-valued function, the
formula in (\ref{lapPower2}) would continue to hold. Second, the Laplacian
$\Delta_{N}$ commutes with applying the trace, so the right-hand side of
(\ref{lapPower2}) is what one obtains by applying the normalized trace to the
right-hand side of (\ref{lapPower1}). Third, the formulas for $\Delta
_{N}(U^{k})$ and $\Delta_{N}(\mathrm{tr}(U^{k}))$ are \textquotedblleft
independent of $N,$\textquotedblright\ meaning that the coefficients of the
various terms on the right-hand side of (\ref{lapPower1}) and (\ref{lapPower2}%
) do not depend on $N.$ This independence holds only because we have chosen to
express things in terms of the \textit{normalized} trace; if we used the
ordinary trace, there would be a factor of $1/N$ in the second term on the
right-hand side of both equations.

Suppose, now, that we wish to apply $\Delta_{N}$ to a product, such as the
function $f(U)=U^{k}\mathrm{tr}(U^{l}).$ As usual with the Laplacian, there is
a product rule that involves three terms, two \textquotedblleft Laplacian
terms\textquotedblright---namely $\Delta_{N}(U^{k})\mathrm{tr}(U^{l})$ and
$U^{k}\Delta_{N}(\mathrm{tr}(U^{l}))$---along with a cross term. The Laplacian
terms can, of course, be computed using (\ref{lapPower1}) and (\ref{lapPower2}%
). The cross term, meanwhile, turns out to be
\[
-\frac{2kl}{N^{2}}U^{k+l}.
\]
Thus, we have%
\[
\Delta_{N}(U^{k}\mathrm{tr}(U^{l}))=\Delta(U^{k})\mathrm{tr}(U^{l}%
)+U^{k}\Delta(\mathrm{tr}(U^{l}))-\frac{2kl}{N^{2}}U^{k+l}.
\]

The behavior in the preceding example turns out to be typical: \textit{The
cross term is always of order} $1/N^{2}.$ Thus, to leading order in $N,$ we
may compute the Laplacian of a function of the form (\ref{tracePoly1}) as the
sum of $n+1$ terms, where each term applies the Laplacian to one of the
factors (using (\ref{lapPower1}) or (\ref{lapPower1})) and leaves the other
factors unchanged.

It should be emphasized that this leading-order behavior applies only if (as
in (\ref{tracePoly1})) we have collected together all of the untraced powers
of $U.$ Thus, for example, if we chose to write $U^{5}$ as $U^{3}U^{2},$ it
would \textit{not} be correct to say that $\Delta_{N}(U^{5})$ is $\Delta
_{N}(U^{3})U^{2}+U^{3}\Delta(U^{2})$ plus a term of order $1/N^{2}.$

The smallness of the cross terms leads to the following \textquotedblleft
asymptotic product rule\textquotedblright\ for the action of $\Delta_{N}$ on
trace polynomials.

\begin{proposition}
[Asymptotic product rule]\label{product.prop}Suppose that $f$ and $g$ are
trace polynomials and that either $f$ or $g$ is \textquotedblleft
scalar,\textquotedblright\ meaning that it contains no untraced powers of $U.$
Then%
\[
\Delta_{N}(fg)=\Delta_{N}(f)g+f\Delta_{N}(g)+O(1/N^{2}),
\]
where $O(1/N^{2})$ denotes a fixed trace polynomial multiplied by $1/N^{2}.$
\end{proposition}

The asymptotic product rule may be interpreted as saying that in the situation
of Proposition \ref{product.prop}, the Laplacian \textit{behaves like a
first-order differential operator}. Furthermore, if, say, $f$ is scalar, then
it turns out that $\Delta_{N}^{n}(f)$ is scalar for all $n,$ which means that
we can apply the asymptotic product rule repeatedly. Thus, by a standard power
series argument, together with some simple estimates (Section 4 of
\cite{DHK}), we conclude that%
\begin{equation}
e^{t\Delta_{N}/2}(fg)=e^{t\Delta_{N}/2}(f)e^{t\Delta_{N}/2}(g)+O(1/N^{2}),
\label{expProduct}%
\end{equation}
assuming at least one of $f$ and $g$ is scalar. The asymptotic product rule,
along with its exponentiated form (\ref{expProduct}), is the key to many of
the results in \cite{DHK}.

Using the asymptotic product rule, along with Proposition \ref{lapPower.prop},
we can readily compute---to leading order in $N$---the Laplacian of any trace
polynomial, as follows.

\begin{proposition}
\label{lap1.prop}For any non-negative integers $k$ and $l_{1},\ldots,l_{M},$
we have%
\begin{align*}
\Delta_{N}(U^{k}\mathrm{tr}(U^{l_{1}})\cdots\mathrm{tr}(U^{l_{M}}))  &
=\Delta_{N}(U^{k})\mathrm{tr}(U^{l_{1}})\cdots\mathrm{tr}(U^{l_{M}})\\
&  +U^{k}\Delta_{N}(\mathrm{tr}(U^{l_{1}}))\mathrm{tr}(U^{l_{2}}%
)\cdots\mathrm{tr}(U^{l_{M}})\\
&  +\cdots\\
&  +U^{k}\mathrm{tr}(U^{l_{1}})\cdots\mathrm{tr}(U^{l_{M-1}})\Delta
_{N}(\mathrm{tr}(U^{l_{M}}))\\
&  +O(1/N^{2}),
\end{align*}
where $O(1/N^{2})$ denotes a fixed trace polynomial multiplied by $1/N^{2}.$
\end{proposition}

Proposition \ref{lap1.prop} leads to a computationally effective procedure for
computing the Laplacian---and therefore also the heat operator---on trace
polynomials, in the large-$N$ case. (See Section 5.1 in \cite{DHK} and Section
8 in \cite{Exposit}.)

\subsection{Concentration properties of the heat kernel measures}

There is one other crucial ingredient needed to understand the large-$N$
limit, namely the concentration properties of the heat kernels on the groups
$U(N)$ and $GL(N;\mathbb{C})$. The concentration properties may be summarized
as saying that the heat kernels are \textit{concentrating onto a singe
conjugacy class} in the limit. Let us consider this at first in the $U(N)$
case. In $U(N),$ a conjugacy class is described by listing the eigenvalues of
the associated matrices. Suppose we choose a matrix $U$ at random from $U(N)$
using the measure $\rho_{t}(U)~dU$ as our probability distribution. Results of
Biane \cite{Biane1}, E. Rains \cite{Rains}, and T. Kemp \cite{Kemp} show that
the eigenvalues of the random matrix $U$ become nonrandom in the limit. To be
more precise, consider for any $U\in U(N)$ the \textbf{empirical eigenvalue
distribution}, which is the probability measure $\gamma_{U}$ on the unit
circle given by%
\[
\gamma_{U}=\frac{1}{N}(\delta_{\lambda_{1}}+\cdots+\delta_{\lambda_{N}}),
\]
where $\lambda_{1},\ldots,\lambda_{N}$ are the eigenvalues of $U.$ The
just-cited results say that there is a certain deterministic measure
$\gamma_{t}$ on $S^{1}$ with the following property: If $U$ is chosen at
random from $U(N)$ using the measure $\rho_{t}(U)~dU,$ then with high
probability when $N$ is large, $\gamma_{U}$ will be close to $\gamma_{t}$ in
the weak sense.

The limiting eigenvalue distribution $\gamma_{t},$ originally identified by
Biane, may be thought of as a deformation of Wigner's semicircular
distribution. That is to say, when $t$ is small, $\gamma_{t}$ has an
approximately semicircular shape in a small neighborhood of 1 in the unit
circle, $S^{1}\cong\lbrack-\pi,\pi).$

Since the eigenvalues of a random matrix $U$ become nonrandom, the normalized
trace of $U$ is also becomes nonrandom in the limit. Specifically,
$\mathrm{tr}(U)$ approaches the value $e^{-t/2},$ in the sense that%
\[
\lim_{N\rightarrow\infty}\left\Vert \mathrm{tr}(U)-e^{-t/2}\right\Vert
_{L^{2}(U(N),\rho_{t})}=0.
\]
This statement means that the heat kernel measure $\rho_{t}(U)~dU$ is
concentrating onto the subset of $U(N)$ where $\mathrm{tr}(U)$ has the value
$e^{-t/2}.$ We have a similar result for any scalar trace polynomial: For for
each $l_{1},\ldots,l_{M}$ and $t>0$ there is a constant $C$ (depending on $t$
and $l_{1},\ldots,l_{M}$) such that
\begin{equation}
\lim_{N\rightarrow\infty}\left\Vert \mathrm{tr}(U^{l_{1}})\cdots
\mathrm{tr}(U^{l_{M}})-C\right\Vert _{L^{2}(U(N),\rho_{t})}=0.
\label{limTracePoly}%
\end{equation}
There is a similar result on the $GL(N;\mathbb{C})$ side, but with all traces
taking the value 1:%
\[
\lim_{N\rightarrow\infty}\left\Vert \mathrm{tr}(Z^{l_{1}})\cdots
\mathrm{tr}(Z^{l_{M}})-1\right\Vert _{L^{2}(GL(N;\mathbb{C}),\rho_{t})}=0.
\]
Thus, all scalar trace polynomials \textit{effectively become constants} when
viewed as elements of $L^{2}(U(N),\rho_{t})$ for large $N,$ and similarly in
$\mathcal{H}L^{2}(GL(N;\mathbb{C}),\mu_{t}).$

It is important to emphasize that the preceding discussion applies only to
\textit{scalar} trace polynomials, but not to those that contain untraced
powers of $U.$ For general trace polynomials, the correct statement is this:
Only the untraced powers of $U$ survive in the limit. That is to say,
\[
\lim_{N\rightarrow\infty}\left\Vert U^{k}\mathrm{tr}(U^{l_{1}})\cdots
\mathrm{tr}(U^{l_{M}})-CU^{k}\right\Vert _{L^{2}(U(N),\rho_{t})}=0,
\]
for all $k,$ where $C$ is the same constant as in (\ref{limTracePoly}).

\subsection{Summary}

Theorem \ref{mainIntro.thm} says that in the large-$N$ limit, the boosted
Segal--Bargmann transform $\mathbf{B}_{t}^{N}$ map a single-variable
polynomial $p$ on $U(N)$ to single-variable polynomial $q_{t}$ on
$GL(N;\mathbb{C}).$ We now summarize the procedure for computing $q_{t},$ in
the case when $p(U)=U^{k}$ is a single power of $U.$

\begin{enumerate}
\item Start with $U^{k}$ and compute $e^{t\Delta_{N}/2}(U^{k}),$ to leading
order in $N.$ Section 5.1 in \cite{DHK} and Section 8 in \cite{Exposit}
describe a recursive procedure for performing this computation. The resulting
function will be a trace polynomial on $U(N).$

\item Holomorphically extend (the leading-order approximation to)
$e^{t\Delta_{N}/2}(U^{k})$ from $U(N)$ to $GL(N;\mathbb{C}).$ This amounts to
replacing the variable $U\in U(N)$ with $Z\in GL(N;\mathbb{C})$ in each trace polynomial.

\item In the resulting trace polynomial on $GL(N;\mathbb{C}),$ evaluate each
factor of $\mathrm{tr}(Z^{l})$ to $1.$ The result will then be a
single-variable polynomial in $Z.$
\end{enumerate}

We illustrate the above procedure in the case $k=3.$ Applying the recursive
procedure in Step 1 gives, to leading order in $N,$
\[
e^{t\Delta_{N}/2}(U^{3})\approx e^{-3t/2}\left\{  U^{3}+t[U\mathrm{tr}%
(U^{2})+2U^{2}\mathrm{tr}(U)]+\frac{3t^{2}}{2}U\mathrm{tr}(U)^{2}\right\}  .
\]
We then replace $U\in U(N)$ with $Z\in GL(N;\mathbb{C}),$ obtaining a trace
polynomial on $GL(N;\mathbb{C}).$ Finally, we evaluate $\mathrm{tr}(Z^{2})$
and $\mathrm{tr}(Z)$ to 1, with the result that%
\[
\mathbf{B}_{t}^{N}(U^{3})\approx e^{-3t/2}\left\{  Z^{3}+t[2Z^{2}%
+Z]+\frac{3t^{2}}{2}Z\right\}  .
\]
Thus, if $p(u)=u^{3}$, the polynomial $q_{t}$ in Theorem \ref{mainIntro.thm}
is given by%
\[
q_{t}(z)=e^{-3t/2}\left\{  z^{3}+t[2z^{2}+z]+\frac{3t^{2}}{2}z\right\}  .
\]

\end{document}